\documentclass[10pt,conference]{IEEEtran}
\usepackage{cite}
\usepackage[table,xcdraw]{xcolor}

\usepackage{hyperref}
\usepackage{multirow}
\usepackage{fontawesome}
\usepackage[many]{tcolorbox} 
\tcbset{
    sharp corners,
    colback = white,
    before skip = 0.2cm,    % add extra space before the box
    after skip = 0.5cm      % add extra space after the box
}                           % setting global options for tcolorbox
\newtcolorbox{boxC}{
    colback = sub, % background color
    boxrule = 0pt,  % no borders
}

\definecolor{main}{HTML}{CFCFCF}    % setting main color to be used
\definecolor{sub}{HTML}{CFCFCF}     % setting sub color to be used

\newcounter{keyTakeAwaysCounter}

\newenvironment{keyTakeAways}[1][Key Take Away]
    {
    \addtocounter{keyTakeAwaysCounter}{1}
        \begin{boxC}
        \faLightbulbO ~ \thekeyTakeAwaysCounter. \textbf{#1}.\\
        }{
        
        \end{boxC}
}

\newcounter{keyRQAnswerCounter}

\newenvironment{keyRQAnswer}[1][RQAnswer]
    {
        \addtocounter{keyRQAnswerCounter}{1}
        \faKey ~ \textbf{#1} \textbf{RQ$_{\thekeyRQAnswerCounter}$}. 
        }{

}

\begin{document}

\title{On Large Language Models in Mission-Critical IT Governance: Are We Ready Yet?}

 \author{\IEEEauthorblockN{Matteo~Esposito}
 \IEEEauthorblockA{\textit{University of Oulu}\\
 Oulu, Finland \\
 matteo.esposito@oulu.fi}
 \and
 \IEEEauthorblockN{Francesco~Palagiano}
 \IEEEauthorblockA{ {\small\textit{Multitel di Lerede Alessandro \& C. s.a.s.}}\\
 Roma, Italy\\
 palagiano.francesco@multitelsrl.it}
 \and
 \IEEEauthorblockN{Valentina~Lenarduzzi}
 \IEEEauthorblockA{\textit{University of Oulu}\\
 Oulu, Finland \\
 valentina.lenarduzzi@oulu.fi}
 \and
 \IEEEauthorblockN{Davide~Taibi}
 \IEEEauthorblockA{\textit{University of Oulu}\\
 Oulu, Finland \\
 davide.taibi@oulu.fi}
 }

\newcommand{\NumberOfResponse}{31 }

\maketitle

\begin{abstract}
\textit{Context}. The security of critical infrastructure has been a pressing concern since the advent of computers and has become even more critical in today’s era of cyber warfare. Protecting mission-critical systems (MCSs), essential for national security, requires swift and robust governance, yet recent events reveal the increasing difficulty of meeting these challenges.

\textit{Aim}.  Building on prior research showcasing the potential of Generative AI (GAI), such as Large Language Models, in enhancing risk analysis, we aim to explore practitioners’ views on integrating GAI into the governance of IT MCSs. Our goal is to provide actionable insights and recommendations for stakeholders, including researchers, practitioners, and policymakers.

\textit{Method.} We designed a survey to collect practical experiences, concerns, and expectations of practitioners who develop and implement security solutions in the context of MCSs. %Analyzing this data will help identify key trends, challenges, and opportunities for introducing GAIs in this niche domain.

\textit{Conclusions and Future Works}. Our findings highlight that the safe use of LLMs in MCS governance requires interdisciplinary collaboration. Researchers should focus on designing regulation-oriented models and focus on accountability; practitioners emphasize data protection and transparency, while policymakers must establish a unified AI framework with global benchmarks to ensure ethical and secure LLMs-based MCS governance.

\end{abstract}


\section{Introduction}
\label{sec:Introduction}
%Since the conception of computers \cite{turing1940mathematical,Turing2009}, security has been a requirement or an issue \cite{esposito2024leveraging}. In the current cyber warfare landscape, where wars are fought over fiber optics cable \cite{esposito2024validate}, protecting critical infrastructure is becoming more daunting and paramount with each passing second \cite{esposito2023can}. Currently, our daily activities almost entirely rely on digital platforms \cite{esposito2024validate}. Anyhow, telecommunication infrastructures enabling surfing the internet for a cooking recipe is not as critical or as fault-tolerant as the one used during a national crisis for military personnel and population coordination \cite{CC2022,PCM_ANS_TI_002,ITSEC1991}. We commonly refer to such vital infrastructure as  ``mission-critical systems'' (MCSs), according to the Common Criteria's definitions \cite{CC2022}. Therefore, the context in which such systems operate is referred to as ``mission-critical context'' (MCC). MCSs need prompt and actionable plans to safeguard and protect them as soon and as much as possible \cite{esposito2024leveraging}. Currently, each MCSs follows standardized rules and security practices that vary among nations and according to the criticality level \cite{CC2022,PCM_ANS_TI_002,ITSEC1991} of the system. Hence, the laws and the standards defining the governance of such systems are fragmented and mostly incompatible among the various nations \cite{it_governance_2023}.
Since the advent of computers~\cite{turing1940mathematical, Turing2009}, security has been a persistent concern~\cite{esposito2024leveraging}. In today’s era of cyber warfare, where conflicts unfold over fiber optic networks~\cite{esposito2024validate}, protecting critical infrastructure has become increasingly vital~\cite{esposito2023can}. While daily activities heavily rely on digital platforms~\cite{esposito2024validate}, telecommunication systems used for casual tasks differ significantly in criticality and fault tolerance from those required for national crises, such as military and population coordination~\cite{CC2022, PCM_ANS_TI_002, ITSEC1991}. These vital systems, known as ``mission-critical systems'' (MCSs)\cite{CC2022}, operate within a ``mission-critical context'' (MCC). MCSs require immediate and robust security measures\cite{esposito2024leveraging}, yet their governance standards remain fragmented and incompatible across nations~\cite{it_governance_2023}.

Governance of MCS (GMCS) involves directing, managing, and controlling IT resources to ensure full compliance with organizational objectives, value maximization, and mitigation of risks associated with core operations in a structured, reproducible method \cite{it_governance_2023,esposito2024leveraging}.  According to the Italian National Security Authority (ANS), GMCS is (\textbf{i}) time-sensitive, (\textbf{ii}) cyber-physically fault-tolerant, and (\textbf{iii}) reproducible. In other words, practitioners design guidelines and enact workflows that are quick and easy to follow (time-sensitive), thoroughly documented (reproducible), and risks are extensively analyzed and managed to minimize them and account for backup plans (cyber-physically fault-tolerant).

Recent events have highlighted that experts often struggle to meet the three core requirements of GMCS~\cite{esposito2024beyond, norris2004mission}. Motivated by these challenges and documented failures~\cite{esposito2024beyond, norris2004mission}, we previously explored the potential of Generative AI (GAI) to assist human experts in performing comprehensive risk analysis, one of the key tasks outlined in ANS and ISO standards~\cite{esposito2024leveraging, esposito2024beyond, iso38500, isaca2019cobit}. Our findings demonstrated that GAI, particularly Large Language Models (LLMs), can support and outperform human experts in speed and accuracy. Therefore, stemming from insights and strong interest from industrial partner regarding the integration of GAI into GMCS~\cite{esposito2024leveraging, esposito2024beyond}, we leveraged  such insights, preliminary designs, ethics, and regulations on the topic, to drive our contributions to the community in future research. Based on those experiences, we tried to study practitioners' views about the usage of GAIs, especially LLMs, inside GMCS. 
More specifically: (\textbf{i}) we provide the first survey on practitioners feelings and perceptions on integrating GAI in GMCS; (\textbf{ii}) we elaborate implications tailored to researchers to guide future works; (\textbf{iii}) we suggest actionable insights for practitioners; (\textbf{iv}) we provide suggestions for policymakers to craft informed regulations and policies.

Our survey highlights the \textbf{essential role of interdisciplinary collaboration} among researchers, practitioners, and policymakers in advancing the safe, ethical, and effective use of LLMs in GMCS. Our findings urge \textbf{Researchers}  to focus on \textbf{regulation-oriented} models, while \textbf{practitioners} highlight the need for \textbf{data protection, transparency, regulatory compliance} and \textbf{accountability} to complement human expertise. Finally, \textbf{policymakers} should work towards a \textbf{unified AI framework}, fostering \textbf{global benchmarks} and \textbf{advisory boards} to ensure that LLMs in GMCS can match the needed security, ethical, and accountability standards.

\textbf{Paper Structure:}  Section~\ref{sec:desing} describes the study design. Section~\ref{sec:Results} presents the obtained results, and Section~\ref{sec:Discussions} discusses them. Section~\ref{sec:Threats} highlights the threats to the validity of our study, and Section \ref{sec:RW} discusses related works. Finally, Section~\ref{sec:Conclusions} draws the conclusion.

\section*{Data Availability Statement}
We publicly provide all raw answers, gathered insights, and questionnaires as part of our replication package, which is hosted on Zenodo\footnote{\url{https://doi.org/10.5281/zenodo.13918422}}.
\section{Study Design}
\label{sec:desing}
This section details the goal and research questions, data collection, and data analysis. Our empirical study follows established guidelines \cite{wholin2012experimentation, DBLP:journals/ese/RunesonH09,Basili1994} and previous similar survey works \cite{rios2020practitioners}.

\subsection{Goal and Research Questions}
The \textit{goal} of our empirical study is to investigate the practitioners perceived benefits, effectiveness, and potential ethical and privacy issues on using LLMs in the governance of mission-critical information technology systems. Our \textit{perspective} is of practitioners seeking to understand the balance between innovation in the decision process and thorough application of regulations in the given context. 
Based on our goal, we defined the following four Research Questions (RQs): 
\begin{boxC}
$RQ_1$. What is the current level of familiarity and experience among MCS practitioners with LLMs?
\end{boxC}
Recent years have witnessed the proliferation of AI-enabled technologies' general availability \cite{akbar20246gsoft}. What was a nice cutting-edge lab experiment is now driving innovation in every possible field \cite{akbar20246gsoft}. Our context is peculiar and dominated by a conservative approach to new technologies. For instance, in our previous investigation in this context \cite{esposito2024leveraging,esposito2024beyond}, we were the first to introduce LLMs in Risk Analysis (RA), a fundamental part of GMCS itself. We highlighted LLMs' role in RA as a valuable assistant to speed up and enhance human experts', i.e., practitioners, capabilities. However, to extend the reach of LLMs capabilities in GMCS and to investigate practitioners' perceptions of the technology and its introduction, they should have a minimal background in using LLMs in MCS. With this first research question, we aim to investigate the current level of familiarity of practitioners with LLMs.

Nonetheless, familiarity is only one face of a multi-faced dice. For instance, innovation is usually blocked by negative experiences or feelings towards a specific technology \cite{diamond1996innovation}. Practitioners exposed to LLMs in MCS for the first time can have different perceived benefits and potential limitations that may allow or prevent the adoption. Hence, we ask:

\begin{boxC}
$RQ_2$. What are the future roles, perceived benefits and potential limitations of LLMs in GMCS?
\end{boxC}

Research is usually faster than industries when the latter are not involved in research and innovation (RI) initiatives \cite{bresciani2021using}. Therefore, researcher-driven innovation can be far from real-world immediate actionability. Hence, research can lead to exciting results that may be hard to implement in a real-world MCC scenario \cite{garousi2020practical}. In a conservative context such as the GMCS, it is paramount to drive innovation slowly and based on the practitioners' perception \cite{laplante2017software}.
In our study, we interviewed practitioners who are currently using LLMs as well as those who are not.
Therefore, this RQ aims to identify potential future roles of LLMs in GMCS as well perceptions of benefits and potential limitations in addressing security challenges.

Practioner's feeling-based research is an increasing trend \cite{saroar2023developers,esposito2023early}. For instance, researchers have investigated early career developers' perceptions of code understandability \cite{esposito2023early}. Similarly, Saroar et al.\cite{saroar2023developers} surveyed perceptions of GitHub actions, while Linberg et al. \cite{linberg1999software} surveyed developer perceptions about software project failure. Moreover, practitioners in GMCS are used to a precise perimeter of action. Therefore, assessing the impact on the established GMCS framework is paramount before introducing a possible game-changing technology. Hence, we ask:

\begin{boxC}
$RQ_3$. What is the impact of integrating LLMs into current in-use GMCS frameworks?
\end{boxC}

GMCSs rely on well-established security protocols \cite{CC2022, PCM_ANS_TI_002, ITSEC1991}. MCC-related standards and regulations are slowly updated and rarely entirely replaced. Hence, the reticence towards integrating new technologies is evident. LLMs have the advantage of quickly processing vast amounts of data \cite{esposito2024beyond} while human obviously fall short \cite{esposito2024leveraging}. Hence, its impact on a consolidated pre-GAI workflow may have disruptive consequences. With this RQ, we aim to grasp practitioners' perception of integrating such technologies into their usual workflow.

Finally, when humanity started organizing into society, it bestowed upon itself rules and regulations \cite{simonton1902origin}. The new society emerging alongside such revolutionizing technologies requires new rules and policies. Hence, we aim to investigate the role of policies and  policymakers  in guiding ethical progress by asking:

\begin{boxC}
$RQ_4$. What is the role of policy frameworks in ensuring LLMs' safe and ethical use in GMCS?
\end{boxC}
We are in the peculiar spot of a fast-changing regulation due to the impact AI is having on society. Recently, primary regulatory bodies, such as the European Union (EU), published a corpus of regulation, namely the AI Act \cite{eu_ai_act_2021}, in which they intended to regulate AI technologies within the perimeter of the EU. It’s one of the first comprehensive efforts by any major regulatory body to address AI's ethical, legal, and safety implications on a large scale. Conversely, the U.S. approach to AI regulation is more fragmented and decentralized compared to the EU’s comprehensive and unified framework \cite{ftc_ai_guidance_2021,algorithmic_accountability_act_2022,national_ai_initiative_act_2020,ostp_ai_bill_of_rights_2022}. While the EU AI Act provides a clear, risk-based regulatory structure, the U.S. has focused on guidelines, principles, and sector-specific regulations rather than a single, overarching AI law. Therefore, in the context of policies and policymakers, we aim to gather insights that can also aid the former in developing or consolidating current frameworks.

% \usepackage{multirow}
% \usepackage{graphicx}
\begin{table*}[ht]
\renewcommand{\arraystretch}{1.3}
\centering
\caption{Survey questions and research questions.\\{\tiny Legend: \textbf{C} - Closed, \textbf{L} - Linkert, \textbf{O} - Open Ended, \textbf{B} - C and O}}
\label{tab:survey}
\scriptsize
\begin{tabular}{crp{0.8\linewidth}c}
\hline
\textbf{RQ}                               & \multicolumn{2}{c}{\textbf{Question}}                                                                                                                                                                      & \textbf{Type} \\ \hline
\multicolumn{1}{l}{\multirow{10}{*}{RQ0}} & \textbf{Q1}                      & How old are you?                                                                                                                                                        & \textbf{C}    \\
\multicolumn{1}{l}{}                      & \textbf{Q2}                      & How do you identify in terms of gender, if you're comfortable sharing?                                                                                                  & \textbf{C}    \\
\multicolumn{1}{l}{}                      & \textbf{Q3}                      & Education Level                                                                                                                                                         & \textbf{C}    \\
\multicolumn{1}{l}{}                      & \textbf{Q4}                      & Geographical Location                                                                                                                                                   & \textbf{C}    \\
\multicolumn{1}{l}{}                      & \textbf{Q5}                      & Organization Size                                                                                                                                                       & \textbf{C}    \\
\multicolumn{1}{l}{}                      & \textbf{Q6}                      & Primary Role in Organization                                                                                                                                            & \textbf{C}    \\
\multicolumn{1}{l}{}                      & \textbf{Q7}                      & Occupation                                                                                                                                                              & \textbf{C}    \\
\multicolumn{1}{l}{}                      & \textbf{Q8}                      & Industry Sector                                                                                                                                                         & \textbf{C}    \\
\multicolumn{1}{l}{}                      & \textbf{Q9}                      & Years of Experience                                                                                                                                                     & \textbf{C}    \\
\multicolumn{1}{l}{}                      & \textbf{Q10}                     & Primary Area of Interest or Expertise                                                                                                                                   & \textbf{C}    \\ \hline
\multirow{3}{*}{RQ1}                      & \textbf{Q11}                     & How familiar are you with large language models (LLMs) such as ChatGPT (GPT-3/4), BERT and similar?                                                                       & \textbf{L}    \\
                                          & \textbf{Q12}                     & Have you ever used LLMs specifically for data or risk analysis (even if not IT-related)?                                                                                & \textbf{C}    \\
                                          & \textbf{Q13}                     & What did you use it for?                                                                                                                                                & \textbf{O}    \\ \hline
\multirow{8}{*}{RQ2}                      & \textbf{Q14}                     & How do you envision LLMs enhancing security risk analysis within mission-critical IT-governance contexts?                                                               & \textbf{C}    \\
                                          & \textbf{Q15}                     & In what specific ways do you envision LLMs enhancing security risk analysis within mission-critical IT-governance contexts?                                             & \textbf{O}    \\
                                          & \textbf{Q16}                     & What are the potential limitations or drawbacks of relying on LLMs for security risk analysis in mission-critical contexts?                                             & \textbf{B}    \\
                                          & \textbf{Q17}                     & What concerns, if any, do you have regarding using LLMs in mission-critical IT governance environments?                                                                 & \textbf{B}    \\
                                          & \textbf{Q18}                     & To what extent do you believe the adoption of LLMs in mission-critical IT governance will improve overall operational efficiency?                                       & \textbf{L}    \\
                                          & \textbf{Q19}                     & To what extent do you believe the adoption of LLMs in mission-critical IT governance will improve overall operational effectiveness?                                    & \textbf{L}    \\
                                          & \textbf{Q20}                     & How confident are you in the ability of LLMs to handle and analyze sensitive or classified information securely?                                                        & \textbf{L}    \\
                                          & \textbf{Q21}                     & How would you rate the current capabilities of LLMs in addressing the specific security challenges faced by mission-critical IT-governance environments?                & \textbf{L}    \\ \hline
\multirow{3}{*}{RQ3}                      & \textbf{Q22}                     & What are your perceptions of integrating LLMs into existing security risk analysis frameworks?                                                                          & \textbf{B}    \\
                                          & \textbf{Q23}                     & Do you agree with the following sentence? 'Integrating LLMs into existing security risk analysis frameworks is easy.'                                                   & \textbf{L}    \\
                                          & \textbf{Q24}                     & Do you agree with the following sentence? 'There are no concerns in integrating LLMs in mission-critical risk analysis.'                                                & \textbf{L}    \\ \hline
\multirow{10}{*}{RQ4}                     & \multicolumn{1}{l}{\textbf{Q25}} & What measures should be implemented to address potential biases or ethical concerns associated with LLM adoption in security risk analysis?                             & \textbf{B}    \\
                                          & \multicolumn{1}{l}{\textbf{Q26}} & How do you perceive the potential impact of LLM adoption on privacy rights and data protection within mission-critical IT governance environments?                      & \textbf{B}    \\
                                          & \multicolumn{1}{l}{\textbf{Q27}} & What regulatory measures are necessary to ensure LLMs' safe and ethical use in mission-critical IT governance?                                                          & \textbf{B}    \\
                                          & \multicolumn{1}{l}{\textbf{Q28}} & How can policies be developed to address the potential risks and challenges associated with the use of LLMs in security risk analysis?                                  & \textbf{B}    \\
                                          & \multicolumn{1}{l}{\textbf{Q29}} & How do you feel about utilizing LLMs to handle sensitive or classified information?                                                                                     & \textbf{O}    \\
                                          & \multicolumn{1}{l}{\textbf{Q30}} & What policies should be implemented to protect data privacy when using LLMs in mission-critical IT environments?                                                        & \textbf{B}    \\
                                          & \multicolumn{1}{l}{\textbf{Q31}} & What standards and guidelines would you recommend to ensure accountability and transparency in using LLMs for security risk analysis?                                   & \textbf{B}    \\
                                          & \multicolumn{1}{l}{\textbf{Q32}} & How can policymakers engage with practitioners and researchers to create a balanced approach to integrating LLMs in mission-critical IT governance?                     & \textbf{B}    \\
                                          & \multicolumn{1}{l}{\textbf{Q33}} & Are you aware of the recent 'AI Act' by the European Union?                                                                                                             & \textbf{C}    \\
                                          & \multicolumn{1}{l}{\textbf{Q34}} & If you're familiar with the recent 'AI Act,' in which category would you place the use of large language models (LLMs) in the context of mission-critical applications? & \textbf{C}    \\ \hline
\end{tabular}
\end{table*}

\subsection{Study Context}
\textbf{Practical Scenario and Previous Insights}. This work stems from our previous experience with domain experts tasked to experimentally and gradually introducing LLMs in the context of GMCS \cite{esposito2024leveraging, esposito2024beyond}. We are in the shoes of an expert tasked with evaluating an IT system in Italy that processes, e.g., a mission critical health information system \cite{esposito2024leveraging}. The evaluation must align with the regulatory requirements of Regulation (EU) 2016/679 (GDPR) and national laws, addressing both general health data protection and the specifics of electronic health records. Additionally, the expert must accommodate the specific needs of the entity using the system.

The expert’s role involves identifying regulatory and business requirements, gathering relevant documentation, and assessing procedures, staff training, security protocols, physical infrastructure, access controls and the system itself. After synthesizing and analyzing this information, the expert conducts a thorough evaluation, prepares a final report, and outlines any necessary action plans.

\textbf{Effective Governance}. According to ISACA \cite{it_governance_isaca}, our work is focused on GMCS, which involves the structured manner of directing, managing, and controlling IT resources to ensure full compliance with organizational objectives, value maximization, and mitigation of risks associated with core operations \cite{it_governance_2023,esposito2024leveraging}. Much emphasis is placed on good governance through oversight, accountability, and decision-making to address IT systems' continued availability, security, and reliability as intrinsic enablers to organizational success. It is that form of governance that is key in ensuring the best performance and compliance with regulatory standards and that the investments being made in technology support the core mission objectives and resilience \cite{it_governance_2023, national_ai_initiative_act_2020,esposito2024beyond}.

Effective governance ensures that IT resources support business goals, facilitates the continuous availability of critical systems, and upholds regulatory requirements. Frameworks like COBIT \cite{isaca2019cobit} and ISO/IEC 38500 \cite{iso38500} are commonly used to structure these processes, helping organizations optimize resource use, mitigate risks, and maintain compliance with industry standards \cite{it_governance_isaca}.

Poor governance causes misalignment, often preventing the proper identification of sensitive data from being compromised by critical services with insufficient security measures \cite{it_governance_isaca, esposito2024beyond}. Poor communication between business and IT leads to poor alignment of priorities and resource allocations, undermining real risk mitigation.

%\subsection{Motivation on the use of LLMs}
\textbf{Motivations for Using LLMs}. Unlike financial and credit risk assessment \cite{mashrur2020machine1, mashrur2020machine2, song2014application}, which primarily involves quantitative data, i.e., such as the number of transactions, their sources and destinations, and their content, RA in our context relies heavily on qualitative aspects \cite{esposito2024beyond}, such as natural language descriptions of scenarios and systems, alongside quantitative data hence the analysis, the anomaly detection and the overall assessment is more challenging. As a result, traditional Machine Learning and other Neural Network models are less suitable for this application compared to LLMs. Despite their limitations, LLMs can significantly enhance data collection and analysis, allowing experts to concentrate on critical tasks like evaluation and action plan development.

\subsection{Population}
We interviewed practitioners and organizations primarily affiliated with the European public administration, specifically Italian ministries, high-ranking military personnel, foreign country embassies, and agencies connected to national security and defense departments and economic partners who insist on the same market. We collected the answers anonymously and provided all participants with information regarding the GDPR act for data collection and protection. We also thoroughly followed the ACM publications policy on research involving human participants and subjects \cite{ACM_Human_Participants_2021}. We leverage our industrial partner, \textbf{Multitel}, an Italian company operating in civil and military security for over 30 years. It is dedicated to researching and developing new technologies for information security and provides products and services aimed at safeguarding data, both at rest and in motion. More specifically, we leverage its network to reach the aforementioned practitioners.

It is noteworthy to consider the peculiar and niche population our investigation targeted. Practitioners who work in mission-critical contexts such as national security, development, and regulating secure communication between ministries and military personnel are usually \textbf{instructed to avoid participating in such research}. We undergo a \textbf{lengthy authorization process}, which has led us to get valuable data regarding 31 practitioners' knowledge and perceptions of using LLMs in such delicate contexts. In the same vein, when answering the question, all the interviewees implicitly \textbf{refer only to on-premise, i.e., private, LLMs}. In our context, no data can exfiltrate the intended geographical location, with granularity ranging from national territory to single offices. Hence, although their reply does not reflect the general audience, they are, on the other hand, a representative sample of the primarily European personnel occupied in such contexts. 

\subsection{Questionnaire}
Table \ref{tab:survey} presents the questions induced by our RQs. Due to space constraints, we provide the complete questionnaire in the replication package. It is worth noticing that, before submitting the questionnaire, we have performed a pilot questionnaire to derive the answers for the closed-ended questions (\textbf{C}) predefined answers. Moreover, for most of the closed-ended questions we provided an open-ended version (\textbf{B}) to account for possible missing categories or to allow room for an in-depth explanation of the selection, though we flag them as optional. We interviewed the experts involved in the previous papers \cite{esposito2024leveraging,esposito2024beyond}, and leveraging domain knowledge, we reached a consensus for the predefined answer to the closed-ended questions. We did not ask the pilot's experts to participate in the final questionnaire to avoid biases.

According to our guidelines \cite{wholin2012experimentation, DBLP:journals/ese/RunesonH09,Basili1994,rios2020practitioners}, we asked the participants to answer demographic questions to extract insights on the population under examination. 

Therefore, Q1 to Q10 refers to the interview's personal background and professional activities. We gave the participants predefined answers to facilitate data analysis by leveraging domain knowledge \cite{esposito2024leveraging}.

For $RQ_1$, we asked the interviewee their degree of familiarity with LLMs and similar technologies. We provided a Linkert scale question (\textbf{L}), Q11, and a closed-ended question, Q12. A Likert scale is a psychometric scale commonly used in questionnaires to gauge respondents’ attitudes, opinions, or perceptions on a particular topic \cite{likert1932technique}. It typically consists of a series of statements where respondents indicate their level of agreement or disagreement on a symmetric agree-disagree scale, usually ranging from ``strongly agree'' to ``strongly disagree.'' This method allows for the measurement of people’s attitudes or feelings toward a subject in a quantitative way. The values for each Linkert Scale question are available in the replication package.

For $RQ_2$, we questioned their confidence in a possible role for LLMs in aiding IT MSCs governance. Questions Q14 and Q16-17 are closed-ended questions. While Q15 is open-ended, we also made room for open-ended answers to motivate the choices for Q16-17 better. Finally, we used the Linkert scale for Q18-21.

Regarding $RQ_3$, we surveyed practitioners' confidence in integrating LLMs in the current GMCS frameworks. We provided a Linkert scale for Q23 and Q24. While we provide a closed-ended and open-ended space for Q22.

Finally, for $RQ_4$, we surveyed the role of policy and regulations in ensuring the safe and ethical use of LLMs in our specific context. All questions in this section had predefined answers in multiple-choice and open-ended spaces for Q25-28 and Q30-32. While Q29 is only open-ended and Q33-34 only closed-ended.

\subsection{Data Analysis}
This section presents the data analysis of our work. Our survey includes closed and open-ended questions. Therefore, we select different analysis methods for the two types of survey output. To analyze the responses to the closed questions, we initially employed descriptive statistics to gain a clearer insight into the data. For ordinal and interval data, we focused on the mode and median to assess central tendency, while for nominal data, we calculated the distribution of participants' choices for each option.

Regarding open-ended questions, we employed qualitative data analysis techniques suggested by Strauss and Corbin  \cite{straus1998techniques} and Seaman and Yuepu \cite{seaman2011measuring}. Qualitative analysis helps answer questions of the form ``What is going on here?'' when we want to learn about what people understand and how they deal with what is happening to them through time and changing circumstances \cite{rios2020practitioners}. Thus, it is an appropriate method for explaining, for example, the practitioners perceived benefits and limitations of LLM in GMCS.

Although we already know that our specific context follows mostly a conservative approach towards new technologies, we do not deem that the responses to RQ2, RQ3, and RQ4 may perfectly align with prior expectations, so we anyway adopted an inductive approach to develop a new theory based on the qualitative data provided. The open-ended questions for RQ2 to RQ4 were manually coded as follows. The first and second authors independently coded responses to related questions (RQ2: Q17, RQ3: Q23, and RQ4: Q24, Q25) using open and axial coding, following Strauss and Corbin~\cite{straus1998techniques}. We addressed disagreement via discussion, and codes were organized into a hierarchy of benefits and limitations until saturation was reached. 
Since the open-ended questions were optional and yielded limited responses, we incorporated relevant answers into the discussion section to support the argument where appropriate.

%Initially, the first and second authors independently coded the responses for subsets of related questions (RQ2: Q17, RQ3: Q23, and RQ4: Q24, Q25). This process began with open coding, as Strauss and Corbin  \cite{straus1998techniques} outlined, followed by axial coding to identify higher-level categories.  Subsequently, the two coders compared their coding and discussed any discrepancies until they reached a consensus. The coding process involved assigning codes to small, coherent units within the responses and organizing the emerging concepts (benefits/limitations) into a hierarchical structure of categories. This process was repeated iteratively until a state of saturation was achieved, meaning no new codes or categories were identified. Anyhow, we remark that open-ended questions were optional; thus, at the end of the coding sessions, due to the lack of answers, hence a lack of many categories worth investigating and citing in the results, we deemed to cite the relevant answers as a part of the argumentation in the discussion section where relevant.

\section{Results}
\label{sec:Results}
This section presents the results of our survey. We collected data from \NumberOfResponse interviewees. In the following, we first summarize the information about the study population before describing the results for each of the RQs. Due to space constraints, we present the data only in a tabular format, avoiding graphs. We note that for the closed-ended questions we decided to mention only the top-three options.
% \usepackage{multirow}
% \usepackage{graphicx}
\begin{table*}[]
\centering
\caption{Participants background information}
\label{tab:demography}
\resizebox{\textwidth}{!}{%
\begin{tabular}{clrlclr}
\cline{1-3} \cline{5-7}
\multirow{6}{*}{Age (\textbf{Q1})}                          & 18-24                               & 6,45\%  &  & \multirow{5}{*}{Occupation (\textbf{Q7})}                             & Consultant                   & 6,45\%  \\
                                                            & 25-34                               & 12,90\% &  &                                                                       & Government Official          & 38,31\% \\
                                                            & 35-44                               & 32,26\% &  &                                                                       & IT/Technology Professional   & 45,16\% \\
                                                            & 45-54                               & 29,03\% &  &                                                                       & Law Enforcement Officer      & 3,23\%  \\
                                                            & 55-64                               & 12,90\% &  &                                                                       & Security Analyst             & 6,45\%  \\ \cline{5-7} 
                                                            & Over 65                             & 6,45\%  &  & \multirow{5}{*}{Industry Sector (\textbf{Q8})}                        & Defense/Security             & 12,90\% \\ \cline{1-3}
\multirow{2}{*}{Gender (\textbf{Q2})}                       & Female                              & 32,26\% &  &                                                                       & Education/Research           & 3,23\%  \\
                                                            & Male                                & 67,74\% &  &                                                                       & Government/Public Sector     & 41,94\% \\ \cline{1-3}
\multirow{4}{*}{Education Level (\textbf{Q3})}              & Bachelor's degree                   & 12,90\% &  &                                                                       & Healthcare/Pharmaceutical    & 3,23\%  \\
                                                            & Doctoral degree                     & 9,68\%  &  &                                                                       & Technology/Software          & 38,71\% \\ \cline{5-7} 
                                                            & High school diploma or equivalent   & 32,26\% &  & \multirow{4}{*}{Years of Experience (\textbf{Q9})}                    & 1-3                          & 12,90\% \\
                                                            & Master's degree                     & 45,16\% &  &                                                                       & 4-6                          & 9,68\%  \\ \cline{1-3}
\multirow{2}{*}{Geographical Location (\textbf{Q4})}        & Europe                              & 83,87\% &  &                                                                       & 7-10                         & 9,68\%  \\
                                                            & North America                       & 16,13\% &  &                                                                       & Over 10                      & 67,74\% \\ \cline{1-3} \cline{5-7} 
\multirow{4}{*}{Organization Size (\textbf{Q5})}            & Enterprise (10.000+ employees)      & 12,90\% &  & \multirow{8}{*}{Primary Area of Interest or Expertise (\textbf{Q10})} & Artificial Intelligence      & 3,23\%  \\
                                                            & Large (501-10,000 employees)        & 16,13\% &  &                                                                       & Building Design and Planning & 3,23\%  \\
                                                            & Medium (51-500 employees)           & 25,81\% &  &                                                                       & Cybersecurity                & 19,35\% \\
                                                            & Small (1-50 employees)              & 19,35\% &  &                                                                       & Data Science/Analytics       & 9,68\%  \\ \cline{1-3}
\multirow{5}{*}{Primary Role in Organization (\textbf{Q6})} & Analyst                             & 6,45\%  &  &                                                                       & Governance/Policy Making     & 35,48\% \\
                                                            & Executive/Managerial                & 32,26\% &  &                                                                       & Risk Management              & 9,68\%  \\
                                                            & Policy Making/Compliance/Regulatory & 32,26\% &  &                                                                       & Software Development         & 3,23\%  \\
                                                            & Researcher                          & 3,23\%  &  &                                                                       & Software Engineering         & 16,13\% \\ \cline{5-7} 
                                                            & Technical Specialist                & 25,81\% &  & \multicolumn{1}{l}{}                                                  &                              &         \\ \cline{1-3}
\end{tabular}%
}
\end{table*}

\subsection{Demographic}
Table \ref{tab:demography} presents the demographics of our population. According to Table \ref{tab:demography}, we note that most of the participants are senior experts with ages ranging from 35 to 64 years old (\textbf{Q1}) with over 10 years of experience in MCC (\textbf{Q9}). It is worth noting that only 45\% of our population have a Master's degree, while 32\% have only a high school diploma (\textbf{Q3}).

Moreover, most interviewees come from Europe, although 16\%  of them come from North America (\textbf{Q4}). Regarding their affiliation, organization size ranged from small companies, usually subsidiaries of bigger ones, for 19\% to enterprises for 12\% of them. On average, $\sim$42\%  of the interviewees belong to  Medium and Large companies or institutions (\textbf{Q5}).

Regarding the interviewees' primary role, most are executive or managerial figures or employed in compliance and regulations roles (\textbf{Q6}). As per the occupation, 45\% of the participants are IT/Technology professionals or government officials (38\%) (\textbf{Q7}). Similarly, most of the interviewees, i.e., $\sim$42\%, are employees of the government or the public administration, such as ministries, embassies, and regional and local governments, while $\sim$39\% of them are in the technology and software sector (\textbf{Q8}).

Finally, regarding the interviewee's primary area of interest or expertise, we note that 35\% of them handle governance and policy-making-related works. At the same time, $\sim$41\% of them are experts in computer science-related fields, spread over Software Development ($\sim$3\%), Software Engineering ($\sim$16\%), Cybersecurity ($\sim$19\%) and Artificial Intelligence ($\sim$3\%)  (\textbf{Q10}).

\subsection{Familiarity and experience with LLMs (RQ$_1$)}
Table \ref{tab:RQ1} shows practitioners' familiarity with LLMs. According to Table \ref{tab:RQ1}, most interviewees are slightly to quite familiar with LLMs, with 16\% being extremely familiar and $\sim$26\% having no experience at all with them. This last results was expected because we were interested both in the practitioners that is currently using LLMs and those who aren't. Moreover, 32\% of the interviewees have specific experiences in employing LLMs for data or risk analysis. 

%Hence, we can affirm that \textbf{most of our interviewees are familiar with general purpose LLMs}. In contrast, a minority of them \textbf{have already employed LLMs for data and risk analysis}.  

\begin{keyRQAnswer}[Take Aways for]
Most of the interviewees are familiar with general-purpose LLMs. Morover, a small percentage of them have already employed LLMs for data and risk analysis.  
\end{keyRQAnswer}

% \usepackage{multirow}
% \usepackage{graphicx}
\begin{table}[htb]
\renewcommand{\arraystretch}{1.5} 
\centering
\scriptsize
\caption{Practitioner's familiarity with LLMs.}
\label{tab:RQ1}
%\resizebox{\linewidth}{!}{%
\begin{tabular}{clr}
\hline
\multirow{5}{*}{LLM Familiarity (\textbf{Q11})}                 & Not at all Familiar & 25,81\% \\
                                                       & Slightly Familiar   & 19,35\% \\
                                                       & Somewhat Familiar   & 12,90\% \\
                                                       & Quite Familiar      & 25,81\% \\
                                                       & Extremely Familiar  & 16,13\% \\ \hline
\multirow{2}{*}{LLMs  for data or risk analysis (\textbf{Q12})} & No                  & 67,74\% \\
                                                       & Yes                 & 32,26\% \\ \hline
\end{tabular}%}
\end{table}

\subsection{Future Roles, Perceived Benefits and Limitations (RQ$_2$)}
Table \ref{tab:RQ2_1} highlights practitioners' perceived benefits, potential limitations, and key concerns surrounding LLMs use in GMCS.  According to  Table \ref{tab:RQ2_1}, practitioners focused on three main future roles for LLMs in RA  (\textbf{Q15}). More specifically, 27\% of the interviewees affirmed that LLMs are useful for automated threat detection and response and enhancing the overall predictive power of RA. Finally, envisioning a continuously running LLM, the interviewee also highlights the potential for real-time anomaly detection. Similarly, practitioners highlighted legal and regulatory compliance challenges, limited contextual understanding, and high computation resource requirements as the top three potential limitations (\textbf{Q16}). Moreover, privacy and data protection, accuracy and reliability of LLMs outputs and potential bias in data analysis were the top three concerns for the practitioners (\textbf{Q17}).

In the same vein, Table \ref{tab:RQ2_2} shows the practitioners' point of view on the efficiency, effectiveness, and confidence of LLMs for IT governance at large. More specifically, according to Table \ref{tab:RQ2_2}, all the interviewees have positive opinions that LLMs will improve operation efficiency  (\textbf{Q18}) as well as the effectiveness  (\textbf{Q19}) of the overall governance. It is worth noticing that, despite the very conservative context in which our study takes place, most participants have somewhat to fairly confidence in having LLMs handling sensitive information (\textbf{Q20}). Similarly, most participants expressed good to excellent confidence in current LLMs capabilities, hence goodness of fit, for improving GMCS (\textbf{Q21}).

%Hence, we can affirm that \textbf{most of our interviewees expressed positive perceived benefits} in having LLMs handling the governance in MCC. Although a generally good opinion \textbf{there are potential limitations and major concerns} that the community needs to address in the future.  Nonetheless, no limitation or concern was prevalent.

\begin{keyRQAnswer}[Take Aways for]
Practitioners expressed positive perceptions on the future roles of LLMs in GMCS. However, legal and regulatory compliance challenges and limited contextual understanding are major concerns our community needs to address in the future.  
\end{keyRQAnswer}

% \usepackage{multirow}
% \usepackage{graphicx}
\begin{table}[]
\renewcommand{\arraystretch}{1.5} 
\centering
\scriptsize
\caption{LLMs' Perceived Benefits and Potential Limitations in GMCS}
\label{tab:RQ2_1}
\resizebox{\linewidth}{!}{%
\begin{tabular}{clr}
\hline
\multirow{5}{*}{Future Role (\textbf{Q14})}          & Automated threat detection and response           & 27,40\% \\
                                            & Enhanced data visualization tools                 & 15,07\% \\
                                            & Enhanced predictive analytics                     & 23,29\% \\
                                            & Improved natural language understanding           & 9,59\%  \\
                                            & Real-time anomaly detection                       & 24,66\% \\ \hline
\multirow{5}{*}{Potential Limitation (\textbf{Q16})} & High computational resource requirements          & 21,88\% \\
                                            & Lack of transparency in decision-making processes & 17,19\% \\
                                            & Legal and regulatory compliance challenges        & 23,44\% \\
                                            & Limited understanding of context and nuance       & 21,88\% \\
                                            & Potential for adversarial attacks                 & 15,63\% \\ \hline
\multirow{5}{*}{Concerns (\textbf{Q17})}            & Accuracy and reliability of LLMs outputs           & 23,08\% \\
                                            & Integration and compatibility issues              & 13,85\% \\
                                            & Potential bias in data analysis                   & 20,00\% \\
                                            & Privacy and data protection concerns              & 27,69\% \\
                                            & Security vulnerabilities                          & 15,38\% \\ \hline
\end{tabular}%
}
\end{table}
% \usepackage{multirow}
% \usepackage{graphicx}
\begin{table}[]
\scriptsize
\renewcommand{\arraystretch}{1.5} 
\centering
\caption{LLM in Mission Critical Governance}
\label{tab:RQ2_2}
\resizebox{\linewidth}{!}{%
\begin{tabular}{cll}
\hline
\multirow{5}{*}{Improve Operational Efficiency (\textbf{Q18})}                                         & No Improvements          & 0\%     \\
                                                                                              & Minor Improvements       & 0\%     \\
                                                                                              & Moderate Improvements    & 38,71\% \\
                                                                                              & Significant Improvements & 35,48\% \\
                                                                                              & Huge Improvements        & 25,81\% \\ \hline
\multirow{5}{*}{Improve Operational Effectiveness (\textbf{Q19})}                                      & No Improvements          & 0\%     \\
                                                                                              & Minor Improvements       & 0\%     \\
                                                                                              & Moderate Improvements    & 35,48\% \\
                                                                                              & Significant Improvements & 45,16\% \\
                                                                                              & Huge Improvements        & 19,35\% \\ \hline
\multirow{5}{*}{Confidence on handling sensitive information (\textbf{Q20})} & Not at all Confident     & 0,00\%  \\
                                                                                              & Slightly Confident       & 16,13\% \\
                                                                                              & Somewhat Confident       & 41,94\% \\
                                                                                              & Fairly Confident         & 38,71\% \\
                                                                                              & Very Confident           & 3,23\%  \\ \hline
\multirow{5}{*}{Confidence on current LLMs capabilities (\textbf{Q21})}                                & Poor     & 0,00\%  \\
                                                                                              & Fair     & 16,13\%  \\
                                                                                              & Good       & 38,71\% \\
                                                                                              & Very Good       & 25,81\% \\
                                                                                              & Excellent         & 19,35\% \\\hline
\end{tabular}%
}
\end{table}

\subsection{Impact on current GMCS frameworks (RQ$_3$)}
Table \ref{tab:RQ3} presents the impact of integrating LLMs in the currently established governance workflow. According to Table \ref{tab:RQ3} nearly 50\% of the interviewees adopted a neutral stance on each question, neither demonizing nor praising LLMs, which reflects their generally conservative approach.

More specifically, 23\%  of the interviewees expressed an optimistic view of the integration, and 46\%  of them stated that LLMs have an innovative potential (\textbf{Q22}). Nonetheless, $\sim$18\% expressed a cautious approach while 5\% a skeptical view (\textbf{Q22}). Furthermore, when asked to agree on the ease of integration (\textbf{Q23}), $\sim$48\% expressed a neutral agreement, and $\sim$23\% expressed a good agreement, with 6\% strongly agreeing and 6\% strongly disagree. Similarly, when asked to agree on the absence of concerns in the integration, (\textbf{Q24})  $\sim$39\% of the participants had a neutral agreement. In contrast, 19\% of them agreed with 6\% strongly agreeing and 12\% strongly disagree.

%Hence, we can affirm that \textbf{most of our interviewees expressed a balanced to positive perception} on the integration of LLMs in the current workflow for the governance in MCC. 
\begin{keyRQAnswer}[Take Aways for]
Practitioners expressed a balanced to positive perception of the integration of LLMs in the current GMCS frameworks. 
\end{keyRQAnswer}

% \usepackage{multirow}
% \usepackage{graphicx}
\begin{table}[]
\centering
\scriptsize
\renewcommand{\arraystretch}{1.5} 
\caption{Integrating LLMs in Current Workflows}
\label{tab:RQ3}
\resizebox{\linewidth}{!}{%
\begin{tabular}{cll}
\hline
\multirow{5}{*}{Perception on Integration (\textbf{Q22})}          & Balanced Perspective   & 7,69\%  \\
                                                          & Cautious Approach      & 17,95\% \\
                                                          & Innovative Potential   & 46,15\% \\
                                                          & Optimistic Integration & 23,08\% \\
                                                          & Skeptical View         & 5,13\%  \\ \hline
\multirow{5}{*}{Ease of Integration (\textbf{Q23})}                & Strongly Disagree      & 6,45\%  \\
                                                          & Disagree               & 16,13\% \\
                                                          & Neutral                & 48,39\% \\
                                                          & Agree                  & 22,58\% \\
                                                          & Strongly Agree         & 6,45\%  \\ \hline
\multirow{5}{*}{Absence of Concerns in Integration (\textbf{Q24})} & Strongly Disagree      & 12,90\% \\
                                                          & Disagree               & 22,58\% \\
                                                          & Neutral                & 38,71\% \\
                                                          & Agree                  & 19,35\% \\
                                                          & Strongly Agree         & 6,45\%  \\ \hline
\end{tabular}%
}
\end{table}

% \usepackage{multirow}
% \usepackage{graphicx}
\begin{table*}[]
\centering
\scriptsize
\renewcommand{\arraystretch}{1.5} 
\caption{LLMs Safety, Ethics and Policies in Mission Critical Governance}
\label{tab:RQ4}
\resizebox{\linewidth}{!}{%
\begin{tabular}{cp{7.3cm}rlcp{8cm}r}
\cline{1-3} \cline{5-7}
\multirow{5}{*}{\parbox{2cm}{\centering How to address biases and ethical issues (\textbf{Q25})}}                        & Creation of industry-wide ethical standards and guidelines                & 23,68\% &  & \multirow{4}{*}{\parbox{2cm}{\centering Policies to be implemented for Data Privacy (\textbf{Q29})}}                                                    & Developing guidelines for secure data storage and transmission                     & 25,37\% \\
                                                                                       & Development of bias detection and mitigation algorithms                   & 19,74\% &  &                                                                                                                       & Enforcing strict data anonymization techniques                                     & 32,84\% \\
                                                                                       & Establishment of diverse and inclusive AI development teams               & 17,11\% &  &                                                                                                                       & Limiting data access to essential personnel only                                   & 20,90\% \\
                                                                                       & Implementation of transparency and accountability measures                & 13,16\% &  &                                                                                                                       & Requiring explicit user consent for data usage                                     & 20,90\% \\ \cline{5-7} 
                                                                                       
                                                                                       & Regular ethics training for AI practitioners                              & 26,32\% &  & 
                                                                                       \multirow{4}{*}{\parbox{2cm}{\centering Initiatives for Promoting Accountability and Transparency (\textbf{Q30})}}                                       & Creating a public registry of LLM applications and their use cases                 & 23,19\% \\ \cline{1-3}
\multirow{5}{*}{\parbox{2cm}{\centering Perceptions on the impact on privacy rights and data protection (\textbf{Q26})}} & Challenges in ensuring data anonymity and confidentiality                 & 23,53\% &  &                                                                                                                       & Developing industry-specific guidelines for LLM usage                              & 27,54\% \\
                                                                                       & Increased risk of data breaches and unauthorized access                   & 19,12\% &  &                                                                                                                       & Defining clear documentation requirements for LLM algorithm and training data & 33,33\% \\
                                                                                       & Legal and regulatory compliance implications                              & 17,65\% &  &                                                                                                                       & Implementing regular third-party audits of LLM systems                             & 15,94\% \\ \cline{5-7} 
                                                                                       & Need for robust encryption and access controls                            & 20,59\% &  & \multirow{5}{*}{\parbox{2cm}{\centering Initiatives for Promoting Policy Makers, Practitioners and Researcher Colaboration on Policies (\textbf{Q31})}} & Encouraging public-private partnerships for collaborative research                 & 27,27\% \\
                                                                                       & Potential erosion of individual privacy rights                            & 19,12\% &  &                                                                                                                       & Establishing advisory boards with representatives from various sectors             & 24,24\% \\ \cline{1-3}
\multirow{4}{*}{\parbox{2cm}{\centering Necessary regulations for Safe and Ethical Use (\textbf{Q27})}}                & Establishing a certification process for LLM technologies                 & 26,32\% &  &                                                                                                                       & Facilitating open forums for discussion and feedback                               & 18,18\% \\
                                                                                       & Implementing strict data privacy laws specifically for LLM applications   & 22,37\% &  &                                                                                                                       & Hosting regular workshops and seminars with stakeholders                           & 28,79\% \\
                                                                                       & Mandating regular audits and compliance checks for LLM deployments        & 28,95\% &  &                                                                                                                       & None of the Above                                                                  & 1,52\%  \\ \cline{5-7} 
                                                                                       & Requiring transparency reports from organizations using LLMs              & 22,37\% &  & \multirow{2}{*}{\parbox{2cm}{\centering EU ``AI Act'' Awareness (\textbf{Q32})}}                                                                           & No                                                                                 & 32,26\% \\ \cline{1-3}
\multirow{5}{*}{\parbox{2cm}{\centering How to develop effective policies for Safety and Ethical Use (\textbf{Q28})}}    & By benchmarking policies from other countries that use LLMs               & 28,17\% &  &                                                                                                                       & Yes                                                                                & 67,74\% \\ \cline{5-7} 
                                                                                       
                                                                                       & By conducting public consultations to gather diverse opinions             & 12,68\% &  & \multirow{3}{*}{\parbox{2cm}{\centering Classification of the approach according to the ``AI Act'' (\textbf{Q33})}}                                       & High-Risk AI systems                                                               & 82,61\% \\
                                                                                       & By focusing solely on technological solutions without additional policies & 2,82\%  &  &                                                                                                                       & Low-Risk AI systems                                                                & 8,70\%  \\
                                                                                       & By forming interdisciplinary committees to draft comprehensive policies   & 29,58\% &  &                                                                                                                       & Prohibited AI systems                                                              & 8,70\%  \\ \cline{5-7} 
                                                                                       & By incentivizing research into the risks of LLMs                          & 26,76\% &  & \multicolumn{1}{l}{}                                                                                                  &                                                                                    &         \\ \cline{1-3}
\end{tabular}%
}
\end{table*}

\subsection{Role of policy in LLM's safe and ethical use (RQ$_4$)}
Table \ref{tab:RQ4} presents the practitioner's concerns and initiatives proposal on safety, ethics, and policies around using LLMs in MCC governance. According to Table \ref{tab:RQ4},
the participants suggest three key actions to address biases and ethical issues: the creation of industry-wide ethical standards and guidelines ($\sim$24\%), the development of bias detection and mitigation algorithms ($\sim$20\%) and regular ethics training for AI practitioners (26\%) (\textbf{Q25}). Moreover, regarding the perceptions of the impact on privacy rights and data protection (\textbf{Q26}), the top three concerns, as highlighted by the practitioners, are challenges in ensuring data anonymity and confidentiality ($\sim$24\%), increased risk of data breaches and unauthorized access (19\%), and need for robust encryption and access controls  ($\sim$21\%).

Regarding the necessary regulations for the safe and ethical use of LLMs in MCC governance (\textbf{Q27}), it is worth noticing that, according to the interviewees, there is no clear winner among the options. The top two options are mandating regular audits and compliance checks for LLM deployments ($\sim$29\%) and establishing a certification process for LLM technologies (22\%). Similarly, regarding methodologies for developing such policies, (\textbf{Q28}), the interviewees suggested forming interdisciplinary committees to draft comprehensive policies  ($\sim$30\%) and benchmarking policies from other countries already implementing LLMs in their MCC governance (28\%).

Regarding data privacy (\textbf{Q29}), the interviewees suggests creating policies enforcing strict data anonymization techniques when LLMs handle sensitive information ($\sim$33\%) and developing national and international guidelines for secure data storage and transmission of trained model data (25\%).

Regarding the promotion of accountability and transparency (\textbf{Q30}), practitioners highlighters mostly define clear documentation requirements for LLM algorithm and training data ($\sim$33\%) and develop industry-specific guidelines for LLM usage ($\sim$28\%).

Futhermore, the interviewees suggested interesting initiatives for Promoting Policy Makers, Practitioners and Researchers' collaboration on policy-making (\textbf{Q31}), such as hosting regular workshops and seminars with stakeholders ($\sim$29\%) and encouraging public-private partnerships for collaborative research (27\%).

Finally, recently, the EU Commission promulgated the ``AI Act''. Most interviewees knew of such recent legislation (\textbf{Q32}) ($\sim$68\%) and classified a possible GMCS handled by LLMs as a high-risk AI system ($\sim$83\%) (\textbf{Q33}).

%Hence, we can affirm that \textbf{practitioners have a clear view of the imminent use of LLMs} in our context. Although a generally positive view, \textbf{only a close collaboration between academia, industry, and legislative bodies} can shape the future of MCC governance while retaining the current level of transparency, safety, and privacy of specialized human agents.

\begin{keyRQAnswer}[Take Aways for]
Most of the interviewees have a clear view of the imminent use of LLMs in our context. Although a generally positive view, only a close collaboration between academia, industry, and legislative bodies can shape the future of GMCS while retaining the needed level of transparency, safety, and privacy according to the current regulations.
\end{keyRQAnswer}
\section{Discussions}
\label{sec:Discussions}
This section discusses our findings and presents implications for researchers, practitioners, and policymakers. Our findings allowed us to grasp the perceptions of practitioners for LLMs in GMCS following the red-thread of our RQs. %Regarding the demographics, we noted that most practitioners interviewed were government-employed young experts from Europe and North America with more than 10 years of experience in the field. 

\textbf{Implications for Researchers}. Our findings suggest that \textbf{researchers have a crucial role} in addressing the limitations and ethical challenges associated with LLM deployment in mission-critical contexts. For instance, in a recent work, Shi et al.~\cite{shi2023large} explored LLMs' \textit{distractibility}, i.e., the tendency for a model to be distracted by irrelevant context. \textbf{Human experts analysing LLMs output }in our past study, highlighted that a small detail in the description of a scenario led the model to hallucinate and make up non-existing threat tied to the small detail \cite{esposito2024beyond} thus confirming Shi et al. work. Therefore, a primary area for further research is the development of context-sensitive LLMs, i.e., \textbf{regulation-oriented} models (\textbf{Q16}).

Our previous attempts showed that RAG improved contextual knowledge, reducing hallucinations in RA \cite{esposito2024beyond}, yet future efforts will be focused on investigating means to realize more regulations-oriented LLMs. For instance, due to the current state of the art, no ISO standards or national and international laws are not algorithmic-friendly regarding how an algorithm, or AI approach, can tackle and operate with zero to no deviation from them. Similarly, LLM-based governance is a branch of research that is not currently receiving substantial research effort. Our work highlighted many aspects, from \textbf{regulation compliance} (\textbf{Q16}) to \textbf{missing ethical framework}, the needs for \textbf{accountable, explainable, and transparent AIs} and its \textbf{reliability} (\textbf{Q17}).

Finally, researchers can contribute by leading \textbf{interdisciplinary collaborations} with industry and policymakers to shape future policies for LLMs in GMCS (\textbf{Q28}, \textbf{Q31}). In leading the research in this field, with close collaboration with practitioners and policymakers on initiatives like regular workshops, audits, and seminars, researchers can contribute to developing and regulating the LLMs-based GMCS.

\begin{keyTakeAways}[Researchers' Roles]
Researchers should focus on frameworks to allow the design of regulation-oriented LLMs. Interdisciplinary collaboration with industry and policymakers, will help shape policies and drive responsible LLM-based governance.
\end{keyTakeAways}

\textbf{Implications for Practitioners}. Regarding practitioners, the adoption of LLMs promises substantial gains in efficiency and operational effectiveness within GMCS, particularly in \textbf{enhancing governance and risk analysis} (\textbf{Q18}). According to \textbf{open-ended answers} to \textbf{Q13}, practitioners are already experimenting with LLMs for \textbf{technical and documentation support}. LLMs are well-suited to automate certain aspects of risk assessment, enabling faster response times and allowing human experts to focus on more strategic aspects of governance as we highlighted before. 
Open-ended answers from \textbf{Q15}, revealed that practitioners grasp LLMs multi-modal analysis capabilities, for ``\textit{rapid identification of possible responses to real-time threat and anomaly detection}'' in terms of ``\textit{overcoming humans limitations}'' and ``\textit{the ability to identify security risks more quickly by simultaneously cross-referencing data from multiple sources}''. While there is optimism surrounding the integration of LLMs, there are significant challenges, especially concerning \textbf{legal and regulatory compliance} and the models’ often limited understanding of complex contexts (\textbf{Q16}, \textbf{Q24}). Therefore, practitioners should leverage LLMs as supportive tools that complement, rather than replace, human expertise.

Open-ended answer to \textbf{Q16} also highlights that the practitioner's unexpected confidence towards LLMs is deemed possible only as long as we do not ``\textit{rely exclusively on it}'' meaning that LLMs need to be mere tools in the hand of human experts, not replacing them. Practitioners evidenced that LLMs ``\textit{might not take into account some variables that are fundamental in the analysis of various problems}''. \textbf{Transparency and accountability} are also key concerns; users should prefer systems that clearly explain how models arrive at their decisions and allow auditing to ensure the systems meet accountability standards (\textbf{Q30}). Those concerns are highlighted in one answer to the open part of \textbf{Q17} that state: ``\textit{LLMs are often considered ‘black boxes,’ making it difficult to understand how they arrive at their conclusions.}'', thus highlighting the daunting need of \textbf{explainable AIs}. In the same vein, a specific practitioner answer to the open-ended part of \textbf{Q22} ``\textit{The introduction of an LLM in a company has the potential to have a great positive impact and it is possible to expect excellent results, but its continuous updating and training is essential to continue to improve it, and not all companies can sustain such an organizational/economic impact}'' summarizes the positive, yet caution approach that is evident when analyzing our collected data. Finally, a \textbf{major impact} emerge as the possibility of having a \textbf{tool} that harmonizes technical requirements, i.e. driven solely by the internal logic of the industry, with regulatory needs. From an industrial perspective, practitioners highlighted the need for a shared tool, perhaps developed by organizations like \textbf{ACM/IEEE}, to \textbf{benchmark software compliance} along with \textbf{ regulatory experts}, both \textbf{national and international}, in the development process of such tools.% This ensures that what is used complies with both domains, providing a more consistent benchmark for evaluating a tool’s validity.

\begin{keyTakeAways}[Key Insights for Practitioners]
Practitioners advocate for strong data protection, transparency, and ethical practices, emphasizing collaboration on policy for shaping the sustainable and responsible future integration of LLMs into GMCS
\end{keyTakeAways}

\textbf{Implications for Policy Makers} Our findings show that policymakers should focus on defining a clear framework for LLMs used in Mission-Critical IT Systems, establishing the rules to keep them current, and monitoring their use, as deduced from Table \ref{tab:RQ4}. The ethical landscape is crucial for practitioners. Supporting regular ethics training and promoting \textbf{collaboration in policy-making} between researchers, policymakers, and industry stakeholders will help foster an environment where all the specific knowledge of each actor involved can be leveraged to craft a state-of-the-art legislation that will guide an ethical and safe integration of GAIs in GMCS (\textbf{Q25}, \textbf{Q31}). Practitioners advocate for standards and regulations that mandate regular audits (\textbf{Q27}), encourage interdisciplinary cooperation, and align with established safety and accountability practices. By actively engaging in these initiatives, practitioners contribute to a future where LLMs can sustainably and ethically augment mission-critical governance processes.  Global AI regulation is fragmented. The EU has its AI Act, while U.S. legislation is still developing. China’s AI law is in draft form \cite{China_AI_Law}, and Russia has an AI Framework running from 2020 to 2025 \cite{Russia_AI_Framework}. The UN adopted a 2024 resolution promoting “safe, secure, and trustworthy” AI, but without a clear global consensus \cite{UN_AI_Resolution}. There is no unified vision or international body to oversee AI development, ethics, and security, leading to inconsistent governance across regions.

Focusing on the perimeter of the EU, it is paramount to accelerate the definition of a more technical framework (\textbf{Q28}) by hosting regular workshops and seminars with stakeholders (\textbf{Q32}), establishing both a permanent public-private partnership for collaborative research (\textbf{Q32}) and an advisory board (\textbf{Q32}), with representatives from various sectors, to respond to any technical (\textbf{Q30} and \textbf{Q31}), security and ethical concerns (\textbf{Q31} and \textbf{Q33}) or questions related to privacy (\textbf{Q26} and \textbf{Q30}) that could arise. 

Experts in our survey and pilot study expressed \textbf{concerns about the limited reasoning capabilities} of LLMs. This perspective aligns with findings by Mirzadeh et al.\cite{mirzadeh2024gsm_symbolic}, who evaluated LLMs’ mathematical reasoning using GSM8K and introduced GSM-Symbolic for more transparent and explainable assessments. Their study revealed significant performance drops when numerical values or question phrasing changed, highlighting the models’ reliance on pattern replication rather than genuine reasoning. These findings, combined with expert feedback, drive our motivation to explore \textbf{Large Reasoning Models} (LRMs), such as OpenAI’s o1 and o1 preview, or open-source alternatives like Marco-o1\cite{zhao2024marcoo1openreasoningmodels}, i.e., model optimized for outputting alongside their response their inner reason, to address this limitation. In the same vein, when clarity and transparency is threatened, experts suggested in the open part of \textbf{Q30} that ``\textit{The use of these tools should result from an accreditation process similar to that required for handling classified security information}''.
Future research and industrial partnerships should establish a foundation for an \textbf{accreditation process} that ensures LLMs transparently disclose their inner reasoning, enabling thorough evaluation and validation as well defining \textbf{strict technical rules and guidelines} (\textbf{Q31}), define precise documentation requirements for LLM algorithm and training data (\textbf{Q27}). Moreover such a collaboration should also develop industry-specific guidelines for LLM usage and establish a certification process for LLM technologies (\textbf{Q27}) and have the powers to enforce regular audits and compliance checks for LLM deployments (\textbf{Q27}). Finally, the creation of interdisciplinary committees to draft comprehensive policies benchmark of policies (\textbf{Q33}) and the benchmarking of policies from other countries that use LLMs is also paramount to keep them up to date and to address them with better solutions adopted elsewhere (\textbf{Q33}).

\begin{keyTakeAways}[Key Suggestions for Policy Makers]
Policymakers should promote a unified AI framework through to set standards and certification processes for LLMs-based products. Benchmarking global policies and fostering interdisciplinary collaboration will ensure secure and verifiable LLMs-based GMSC.
\end{keyTakeAways}

%\begin{keyTakeAways}[Key Insight and Action Point]
%Lawmakers need a unified regulatory framework for mission-critical LLMs, with clear guidelines on monitoring, security, and ethics to address global regulatory fragmentation.
%Thus, they should form an interdisciplinary advisory board to create technical guidelines, certification processes, and enforcement mechanisms for LLMs, while benchmarking global best practices.
%\end{keyTakeAways}

\section{Threats to Validity}
\label{sec:Threats}
In this section, we discuss the threats to the validity of our case study. We categorized the threats in Construct,  Internal, External, and Conclusion validity following established guidelines \cite{wholin2012experimentation} and previous similar survey works \cite{rios2020practitioners}.

\textbf{Construct Validity} concerns how our measurements reflect what we claim to measure \cite{wholin2012experimentation}. Our design choices, measurement process, and data filtering may impact our results. To address this threat, we based our choice on past studies and well-established guidelines in designing our methodology \cite{DBLP:journals/ese/RunesonH09,Basili1994}. Moreover, threats to the construct validity may arise due to the behavior of both participants and researchers conducting the study. The participants may act differently merely because they know they are being studied \cite{wholin2012experimentation}. The invitation email mentioned the aim of the study to avoid hypothesis guessing and evaluation apprehension risks. The subjects were also invited to reply in light of their personal experiences. They had been assured that the questionnaire was anonymous and the identity of the persons would not be considered while considering the data.

On the issue of the researcher's expectations, which may involve bias either wittingly or unwittingly due to the expected outcome of the research, we used several researchers who would act as both internal and external reviewers of the questionnaire. This reduced this potential threat to a minimum.

\textbf{Internal Validity} is the extent to which an experimental design accurately identifies a cause-and-effect relationship between variables \cite{wholin2012experimentation}. \textit{Maturation} and \textit{instrumentation} pose significant threats to the internal validity of this study. 

\textit{Instrumentation} refers to the potential impact of the tools or materials used during the study, in this case, the questionnaire. A poorly designed questionnaire can compromise the reliability of the study’s results \cite{wholin2012experimentation}.
To address this threat, we designed the questionnaire with only direct questions, minimizing the need for interpretation and reducing the risk of misunderstandings that could lead to irrelevant answers. Additionally, the questionnaire underwent several rounds of validation (two internal and one external) and a pilot test to identify any inconsistencies or potential misunderstandings before the survey was conducted.

\textit{Maturation} refers to the possibility that participants’ responses may change over time, mainly if the questionnaire is too lengthy \cite{wholin2012experimentation}. Therefore, we designed the questionnaire to be completed within a reasonable timeframe. During the pilot study, we found an average completion time of about 15 minutes. An early indication that this threat was minimized is that all participants finished the survey. Even though the questionnaire included many open-ended questions, the responses were detailed and thoughtful, indicating that participants remained engaged throughout the survey. Participants who did not answer specific questions were consistent with their prior responses, suggesting a clear understanding of the content.

\textbf{External Validity} concerns how the research elements (subjects, artifacts) represent actual elements \cite{wholin2012experimentation}. The sample size is relatively small compared to studies on the same topic but is justified by the niche context of GMCS practitioners. To address this limitation we increased the diversity of participants by varying their organizational size, expertise, experience, and geographical location. Since the laws and regulations considered in our study is mostly international, our findings is relevant for other countries within the bounds of such laws and standards. Future works will focus on collaborating with overseas partners to carry out empirical studies leading to a deeper understanding of the potential and limitations of LLMS in GMCS. Moreover lack of control in selecting the samples constitute a threat, which might introduce unbalanced bias from participants with strong opinions. Nevertheless, we found that a considerable proportion of participants were sitting on the fence, with a cautious positive perceptions, so the probability of a ``strong opinions'' bias is probably slight.
%Our sample size, though not as large as another similar study, is justified by the niche context the company operates in and is still smaller than those reflecting all possible contexts of IT MCS practitioners. We tried to mitigate this threat by increasing the diversity of the participants in organizational size, expertise, and experience and geographical location.  Nonetheless, due to the referenced process and regulations being primarily international, our findings may represent other countries, though considering the implementation allowance the international laws grant to each nation. Given these limitations, we plan to survey with overseas partners to synthesize results across time and the development of a more robust, empirically supported understanding.  
%Another potential threat is the lack of control over the sample selection, which may cause the sample to be biased toward individuals who are very interested or biased toward the subject under study. This may, hence, outweigh the positive or negative perception results. Nevertheless, we found that a considerable proportion of participants at baseline mainly had cautious positive perceptions, so any negative bias is probably slight.

\textbf{Conclusion Validity} focuses on how we draw conclusions based on the design of the case study, methodology, and observed results \cite{wholin2012experimentation}. 
In the context of a survey, the main threat to conclusion validity originates in the coding activity itself, which is variable as a human creative process \cite{rios2020practitioners}. As discussed in the questionnaire description, due to a lack of answers, we did not proceed with coding and cited the relevant answers directly as supporting arguments in the discussions. % To mitigate this, we prepared to implement several steps. The first was a pilot phase of coding when consensus could be reached about the first list of codes (so that, related to each code, there was a shared understanding about what the code referred to and the wording and level of abstraction were shared), the two researchers coded independently, then, they discussed their results.
%Moreover, standardization of terms would have been applied in some activities regarding the coding process, which is also a subjective activity that is prone to cause bias. Again, we applied the strategy above. 
\section{Related Works}
\label{sec:RW}
Since the general availability of GAI models, most prominently, LLMs, consumers, practitioners, and researchers have directed their interest toward this 'new' technology\cite{esposito2024beyond}.

The scientific landscape is witnessing a surge in research on the impact of LLMs across various domains, from healthcare to software development and security \cite{esposito2024beyond, esposito2024leveraging, 10449667, 10.1145/3626252.3630927, 10.1145/3639476.3639764}. These models, treated as “copilots of the innovation” \cite{esposito2024beyond}, represent a pinnacle of research innovation, offering significant potential to advance both academic and industrial applications \cite{10449667, esposito2024leveraging}.

For instance, Fan et al. \cite{10449667}, explored the emerging use of LLMs in Software Engineering, identifying both their potential for enhancing creativity and novelty in SE tasks and the technical challenges, such as filtering incorrect solutions. They emphasized the importance of hybrid approaches that integrate traditional SE methods with LLMs to develop reliable and effective solutions.

In the same vein as the general cautious approach towards these new advances, Sallou et al. \cite{10.1145/3639476.3639764}  explored LLM applications in SE tasks and raised concerns about validity threats, including data leakage and reproducibility. They proposed guidelines for researchers and LLM providers, illustrating their use with a test case generation example.

Differently from past studies we focused on a peculiar context. We recently  tackled a primer for LLMs in MCs RA \cite{esposito2024beyond,esposito2024leveraging}, building a cornerstone on which fellow researchers and industrial practitioners can build. To deepen our investigation, we designed the present survey to grasp practitioners' perceived benefits and fears in the adoption at large in IT MCS governance.

%The usual perception of new or trending technologies is negative or, worst case, desk-rejected.  
\section{Conclusions}
\label{sec:Conclusions}
This section concludes our work. With our survey, we aimed to identify the perceived benefits and limitations, and the impact that LLMs may have on the current GMCS frameworks

%Our survey highlights that \textbf{LLMs are seen as valuable tools for enhancing efficiency and predictive capabilities in mission-critical IT governance}. Practitioners acknowledge their potential to streamline MCC IT governance. However, they also recognize significant challenges, including \textbf{privacy concerns, the need for contextual accuracy, and ethical issues} such as bias and transparency.
%As LLMs are further integrated into sensitive applications, practitioners emphasize the importance of \textbf{balanced adoption}, focusing on responsible and ethical use. This requires both developers and end users to address \textbf{security and compliance} proactively, ensuring that LLMs are deployed with robust data protection measures and align with industry standards. Furthermore, there is a clear call for \textbf{collaborative efforts across industry, policy makers, and academia} to develop guidelines and policies that support the safe and transparent use of LLMs in mission-critical settings.

%
Our survey highlights the \textbf{essential role of interdisciplinary collaboration} among researchers, practitioners, and policymakers in advancing the safe, ethical, and effective use of LLMs in GMCS. \textbf{Researchers} are urged to focus on \textbf{regulation-oriented} models and \textbf{accountability}, while \textbf{practitioners} highlight the need for \textbf{data protection, transparency, and regulatory compliance} to complement human expertise. Finally, \textbf{policymakers} should work towards a \textbf{unified AI framework}, fostering \textbf{global benchmarks} and \textbf{advisory boards} to ensure that LLMs in GMCS can match the needed security, ethical, and accountability standards. Our future research efforts will leverage the insights of our survey as the foundation to design how to deliver to the community what practitioners perceive as necessary for the LLM-based GMCS to come to fruition.

\section*{Acknowledgment}
This work has been partially funded by projects Business Finland 6GSoft,  and Research Council of Finland MuFAno (grants n. 349487 and 349488).

\bibliographystyle{IEEEtran}
\bibliography{main}
\end{document}